\documentclass[12pt]{article}
\usepackage{pic02}
\usepackage{hyperref}
\usepackage{url}
\usepackage{epsfig}

\newcommand{\beight}{\mbox{$^8$B}}

\newcommand{\thetaz}{\mbox{$\theta_z$}}

\newcommand{\cosz}{\mbox{$\cos \theta_z$}}

\newcommand{\hep}{\mbox{{\it hep}}}

\newcommand{\snoccfluxshort}{1.76^{+0.06}_{-0.05}\mbox{(stat.)}^{+0.09}_{-0.09}~\mbox{(syst.)}} 
\newcommand{\snoesfluxshort}{2.39^{+0.24}_{-0.23}\mbox{(stat.)}^{+0.12}_{-0.12}~\mbox{(syst.)}} 
\newcommand{\snoncfluxshort}{5.09^{+0.44}_{-0.43}\mbox{(stat.)}^{+0.46}_{-0.43}~\mbox{(syst.)}} 
 
\newcommand{\snomutauflux}{3.41^{+0.45}_{-0.45}\mbox{(stat.)}^{+0.48}_{-0.45}~\mbox{(syst.)}} 
\newcommand{\snoeflux}{1.76^{+0.05}_{-0.05}\mbox{(stat.)}^{+0.09}_{-0.09}~\mbox{(syst.)}} 
 
\newcommand{\snomutaufluxcomb}{3.41^{+0.66}_{-0.64}}

\newcommand{\nsigmassno}{5.3}

\newcommand{\text} {\rm}
\begin{document}

\title{\bf SOLAR NEUTRINOS}
\author{A.B. McDonald \\
{\em SNO Institute, Queen's University, Kingston, Ontario, Canada}}
\maketitle

%
% photograph of author
%  This is where we will insert a photograph. To see what it would look like,
%  uncomment the following lines.
%
%\begin{figure}[h]
%\begin{center}
%
% include photograph for proceeding version
%
%\includegraphics[height=4.5cm]{art_1.eps}
%
% insert a fixed vertical spacing instead for the ArXiv preprint
%
%\vspace{4.5cm}
%
%\end{center}
%\end{figure}

\baselineskip=14.5pt
\begin{abstract}
Present results and future measurements
 of solar neutrinos are
 discussed. The results to date indicate that solar
 electron neutrinos are changing to other active types and that
 transitions solely to sterile neutrinos are disfavored.
 The flux of $^{8}B$ solar neutrinos produced in the Sun,
 inferred assuming only active neutrino types, is found to be
 in very good agreement with solar model calculations.
  Future measurements will focus on greater accuracy for
 charged current and neutral current sensitive reactions
 to provide more accurate measurements of neutrino flavour
 change and further studies of day-night flux differences
 and spectral shape. Other experiments sensitive to lower energy
 solar neutrinos will be in operation soon. 
   
\end{abstract}
\newpage

\baselineskip=17pt

\section{Present Status of Solar Neutrino Experiments}

Starting the 1960's with the pioneering experiments of Davis and his collaborators \cite{bib:homestake}
using Chlorine as a solar neutrino detection medium, a
 discrepancy was identified between the experimental
measurements and the theoretical calculations \cite{bib:bpb,TC} for solar neutrino fluxes.
 Table 1 lists the results
 for neutrino fluxes from experiments up to the year 2000 and compares them
 with solar model calculations. The rates are factors of two or three lower than predictions
 in each case, leading to the conclusion that either solar models are incomplete or 
there are processes occurring such as flavor change to neutrino types for which
 the experiments have little or no sensitivity. This 30-year old discrepancy had come
to be known as the "Solar Neutrino Problem".

Many attempts have been made to understand these discrepancies in terms of modifications to the solar
model, without significant success. The results may be understood in terms of neutrino flavor change in
vacuum or with matter enhancement in the sun\cite{MSW}.  However, clear interpretation 
in terms of neutrino flavor change depends upon solar models because the various experiments have different
thresholds and are sensitive to different combinations of the nuclear reactions in the sun.
Therefore solar model-independent
approaches have also been pursued experimentally to seek an unambiguous indication of flavor change. These
approaches have included searches for spectral
distortion, day-night and seasonal flux differences by Super-Kamiokande\cite{bib:superk1,bib:superk}.
 These measurements
are accurate but have provided no clear indication of flavor change to date. 

\begin{figure} [!ht]
\begin{center}
\includegraphics[width=8cm,clip]{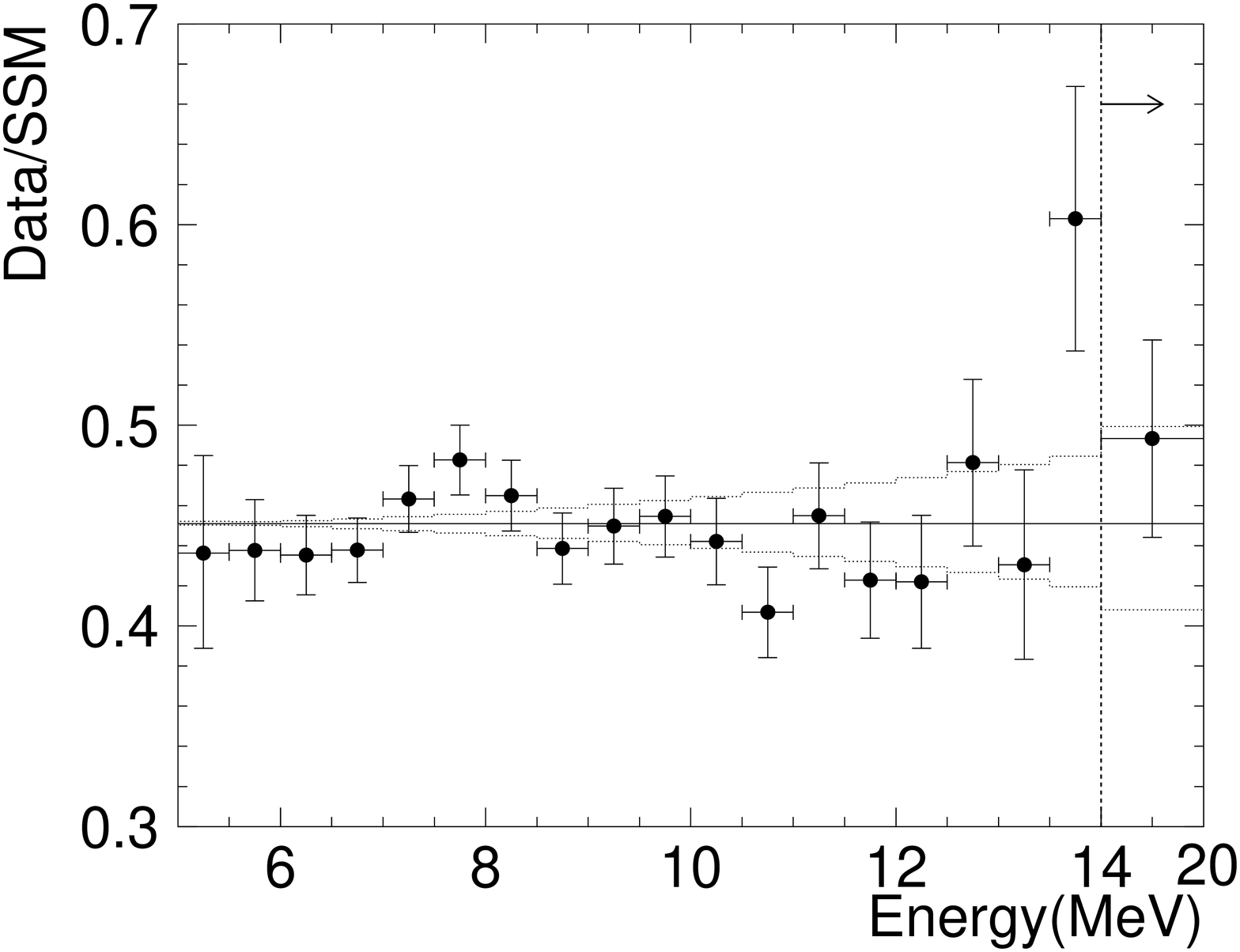}
\end{center}
\caption{The measured \beight\ + \hep\ solar neutrino spectrum from Super-Kamiokande~\cite{bib:superk1}
relative to that of Ortiz {\it et al.} normalized to the SSM~\cite{bib:bpb}.  The data
from 14~MeV to 20~MeV are combined into a single bin.  The horizontal
solid line shows the measured total flux, while the dotted band around
this line indicates the energy correlated uncertainty.  Error bars
show statistical and energy-uncorrelated errors added in quadrature.}
\label{spec}
\end{figure}

The possible effect of neutrino flavor regeneration in the earth after flavor change in the Sun has been
studied through the day-night asymmetry, defined as ${\cal A} = (\Phi_n -
\Phi_d)/\Phi_{average}$, where $\Phi_{average} = \frac{1}{2} (\Phi_n + 
\Phi_d)$.  The Super-Kamiokande results for $^{8}B$ neutrinos \cite{bib:superk} are:
\begin{displaymath}
{\cal A} = 0.021 \pm 0.021~(\rm{stat.})^{+0.013}_{-0.012}~(\rm{sys.})
\end{displaymath}
showing no clear indication of regeneration. Figure 1 shows the dependence of the solar neutrino flux
measured by Super-Kamiokande~\cite{bib:superk1} on zenith angle as well as the day-night comparison.
 Studies by Super-Kamiokande of
seasonal variations also show no evidence for effects other than detected rate changes arising from
the eccentricity of the earth's orbit\cite{bib:superk}.

\begin{figure} [!ht]
\begin{center}
\includegraphics[width=8cm,clip]{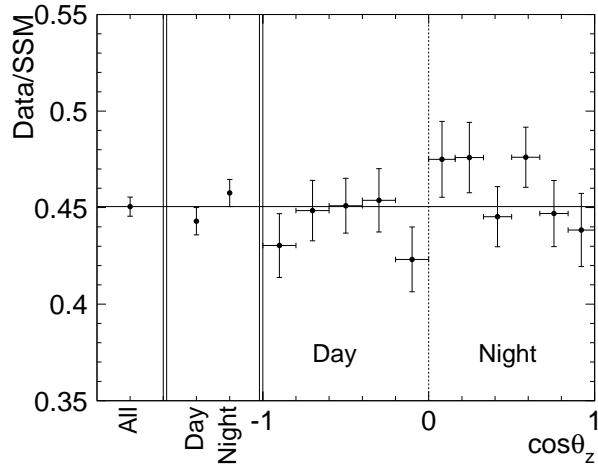}
\end{center}
\caption{The solar zenith angle (\thetaz) dependence of the solar
neutrino flux measured by Super-Kamiokande(error bars show statistical error).  The width of the
night-time bins was chosen to separate solar neutrinos that pass
through the Earth's dense core ($\cosz \ge 0.84$) from those that pass
through the mantle ($0 < \cosz < 0.84$).  
The horizontal line shows the flux for all data.}
\label{dn}
\end{figure}
 
\begin{table*}[htp]
\caption{Summary of solar neutrino observations at different 
solar neutrino detectors. Homestake, SAGE, Gallex, and GNO fluxes are quoted in units of SNU.
The Kamiokande and Super-Kamiokande flux measurements are quoted in units of 10$^{6}$~cm$^{-2}$~s$^{-1}$.}

\begin{tabular}{lllll} \hline
	Experiment & Measured Flux & SSM Flux~\cite{bib:bpb} & Ref. \\ \hline
	Homestake  &  
	2.56$\pm$0.16(stat.)$\pm$0.16(sys.) & 7.6$^{+1.3}_{-1.1}$& 
	\cite{bib:homestake} \\ \hline
	SAGE  & 
	70.8$^{+5.3}_{-5.2}$(stat.+sys.)  & 128$^{+9}_{-7}$& \cite{bib:sage} 
	\\ \hline
	Gallex & 
	77.5$\pm$6.2(stat.)$\pm$4.5(sys.)  & 128$^{+9}_{-7}$ & \cite{bib:gallex} 
	\\ 
	GNO & 
	65.2$\pm$6.4(stat.)$\pm$3.0(sys.) & 128$^{+9}_{-7}$ & \cite{bib:gno} 	
	\\ \hline
	Kamiokande & 	
	2.80$\pm$0.19(stat.)$\pm$0.33(sys.)& 
	5.05$\left(1^{+0.20}_{-0.16}\right)$& \cite{bib:kamioka} \\ 
	Super-Kamiokande & 	
	2.35$\pm$ 0.025(stat.)$^{+0.07}_{-0.06}$(sys.) & 
	5.05$\left(1^{+0.20}_{-0.16}\right)$& \cite{bib:superk} \\ \hline

\end{tabular}\\[2pt]
\protect\label{tbl:solarnuexp} 

\end{table*}

The recent measurements\cite{bib:snoprl2,bib:snoprl1} by the Sudbury Neutrino Observatory (SNO) of interactions 
of $^{8}B$ solar neutrinos in a heavy water detector provide a solar-model-independent measurement
 of neutrino flavor change by comparing charged current (CC) and neutral current (NC) interactions
 with deuterium. The Charged Current (CC) reaction 
\begin{equation}
       d + \nu_{e} \rightarrow p + p + e^{-}  \nonumber
\end{equation}
is specific to electron neutrinos, whereas the Neutral Currrent (NC) Reaction:
\begin{equation}
        \nu_{x} + d \rightarrow n + p + \nu_{x}  \nonumber
\end{equation}
is sensitive to all non-sterile neutrino types equally.

The Elastic Scattering (ES) reaction on electrons: 
\begin{equation}
        \nu_{x} + e^{-} \rightarrow e^{-} + \nu_{x}  \nonumber
\end{equation}
is also sensitive to non-sterile neutrino types other than electron neutrinos,
although they have about six times smaller
cross section than electron neutrinos. Therefore, another solar-model-independent
measurement of neutrino flavor change can be obtained by comparing the CC interactions
with the ES interactions.  

The SNO collaboration have recently reported results from 306 live days of data with pure heavy water
as a detection medium\cite{bib:snoprl2}. The flux of $\nu_e$'s from $^8$B decay is measured by 
the CC reaction rate, assuming no spectral distortion. Comparison of $\phi^{\rm CC}(\nu_e)$ to the
value of $\phi^{\rm NC}(\nu_x)$ provides a null hypothesis test for neutrino flavor change.
The flux of active neutrinos or anti-neutrinos other than electron neutrinos inferred from
these two measurements yields a $5.3\sigma$ difference,
assuming the systematic uncertainties are normally distributed,
providing clear evidence that there is a non-electron neutrino flavor active
component in the solar flux.  

\begin{figure}[!htb]
\begin{center}
\epsfig{figure=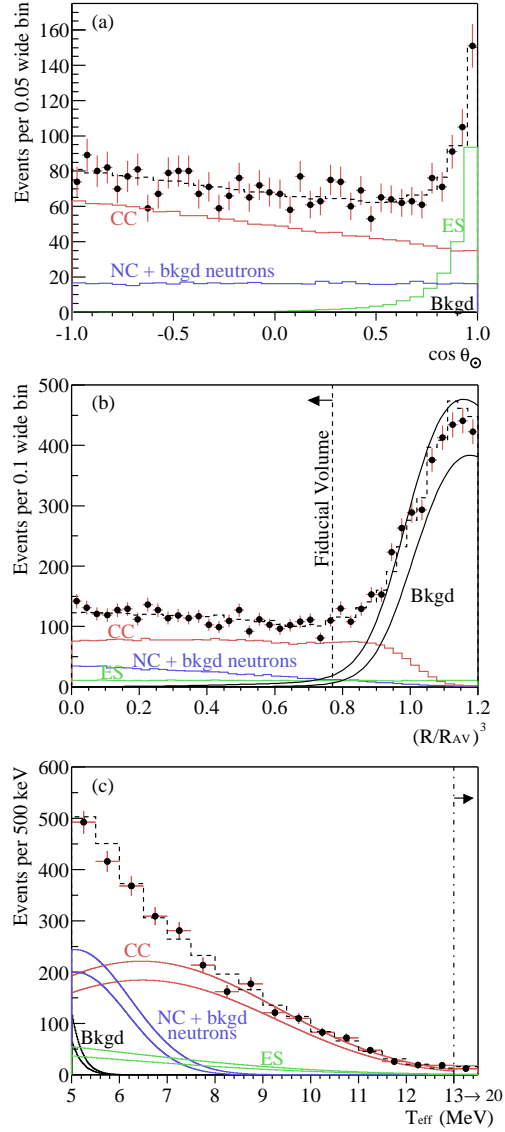,height=6.0in}
\end{center}
\caption{\label{SNOfig2}(a) Distribution of $\cos\theta_{\odot}$ for $R \le 550$ cm.
 (b) Distribution of the volume weighted radial variable $(R/R_{\rm AV})^{3}$.
  (c) Kinetic energy for $R \le 550$ cm.  Also shown are the Monte Carlo predictions for CC,
 ES and NC + bkgd neutron events scaled to the fit results, and the calculated spectrum
 of Cherenkov background (Bkgd) events.  The dashed lines represent the summed components,
 and the bands show $\pm 1\sigma$ uncertainties.  All distributions are for events with
 $T_{\rm eff}$$\geq$5 MeV. }

\end{figure}

Figure~\ref{SNOfig2} (a) displays the
distribution obtained by SNO for $\cos\theta_\odot$,  the angle
between the reconstructed direction of the event and the instantaneous
direction from the Sun to the Earth.  The forward peak in this
distribution arises from the kinematics of the ES reaction, while
CC electrons are expected to have a distribution which is 
$(1-0.340\cos\theta_\odot)$~\cite{vogel}, before accounting for detector 
response.  

Normalized to the integrated rates above the kinetic energy threshold
of $T_{\rm eff}$$\geq 5$~MeV, the  flux of $^8$B neutrinos measured
with each reaction in SNO, 
assuming the standard spectrum shape~\cite{ortiz} is (all fluxes are presented in
units of $10^6~{\rm cm}^{-2} {\rm s}^{-1}$):
\nobreak{
\begin{eqnarray*}
\phi^{\text{SNO}}_{\text{CC}} & = & \snoccfluxshort \\
\phi^{\text{SNO}}_{\text{ES}} & = & \snoesfluxshort \\
\phi^{\text{SNO}}_{\text{NC}} & = & \snoncfluxshort. 
\end{eqnarray*}}
\noindent Electron neutrino cross sections are used to calculate all fluxes.
The excess of the NC flux over the CC and ES fluxes implies neutrino flavor transformations.
The result for the total active neutrino flux obtained with the NC reaction is in very good
agreement with the value calculated~\cite{bib:bpb}
by solar models: $5.05\pm 1.0 \times 10^6~{\rm cm}^{-2} {\rm s}^{-1}$.

A simple change of variables resolves the data directly into electron ($\phi_{e}$) and
 non-electron ($\phi_{\mu\tau}$)
 components. This change of variables allows a direct test of the null hypothesis
 of no flavor transformation ($\phi_{\mu\tau}=0$) without requiring calculation of the CC, ES, and NC
 signal correlations.  
\begin{eqnarray*}
\phi_{e} & = & \snoeflux \\
\phi_{\mu\tau} & = & \snomutauflux 
\end{eqnarray*}
\noindent assuming the standard ${}^{8}$B shape.
Combining the statistical and systematic uncertainties in quadrature,
 $\phi_{\mu\tau}$ is $\snomutaufluxcomb$,
which is \nsigmassno$\sigma$  above zero, providing  strong
evidence for flavor transformation consistent with neutrino oscillations~\cite{maki,pontecorvo}.

\begin{figure}
\includegraphics[width=3.7in]{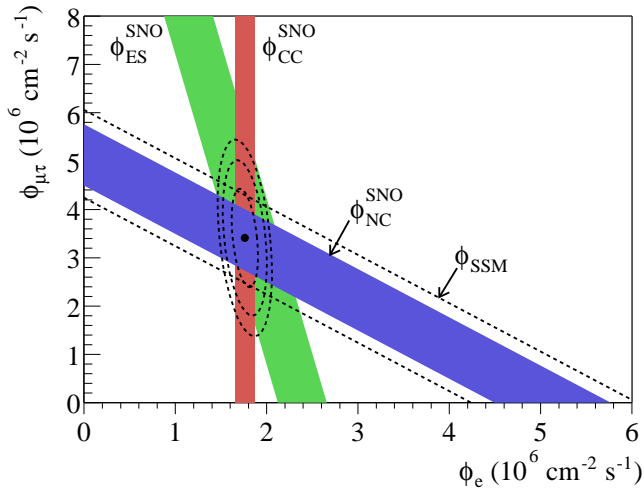}
\caption{\label{hime_plot}Flux of ${}^{8}$B solar neutrinos which are $\mu$ or $\tau$ flavor
 vs flux of electron neutrinos deduced from the three neutrino reactions in SNO.  The diagonal
 bands show the total ${}^{8}$B flux as predicted by the SSM~\cite{bib:bpb} (dashed lines) and that
 measured with the NC reaction in SNO (solid band).  The intercepts of these bands with the axes
 represent the $\pm 1\sigma$ errors.  The bands intersect at the fit values
 for $\phi_{e}$ and $\phi_{\mu\tau}$, indicating that the combined flux results are consistent
 with neutrino flavor transformation assuming no distortion in the ${}^{8}$B neutrino energy spectrum.}
\end{figure}

Figure~\ref{hime_plot} shows the flux of non-electron flavor active neutrinos
vs the flux of electron neutrinos deduced from the SNO data.  The three bands
 represent the one standard deviation measurements of the CC, ES, and NC rates.
 The error ellipses represent  the 68\%, 95\%, and 99\% joint probability contours
 for $\phi_{e}$ and $\phi_{\mu\tau}$.

Spectra for day and night time periods have also been obtained by SNO that show no statistically significant
differences, in agreement with the previous measurements by Super-Kamiokande. If oscillation solely to a sterile
neutrino is occurring,
the SNO CC-derived $^8$B flux above a threshold of 6.75 MeV will be 
consistent with the integrated Super-Kamiokande ES-derived 
$^8$B flux above a threshold of 8.5 MeV\cite{lisi1}. Adjusting the ES 
threshold\cite{bib:superk1} this derived flux difference is $0.53\pm 0.17 \times 10^{6}$ cm$^{-2}{\rm s}^{-1}$, or 
3.1$\sigma$ away from zero, implying that the oscillation is not solely to sterile neutrinos. More recent results
SNO~\cite{bib:snoprl2} and Super-Kamiokande~\cite{bib:superk} with lower thresholds increase this difference to more than
4.5$\sigma$ away from zero.

\begin{figure*}
\begin{center}
\includegraphics[width=3.5in]{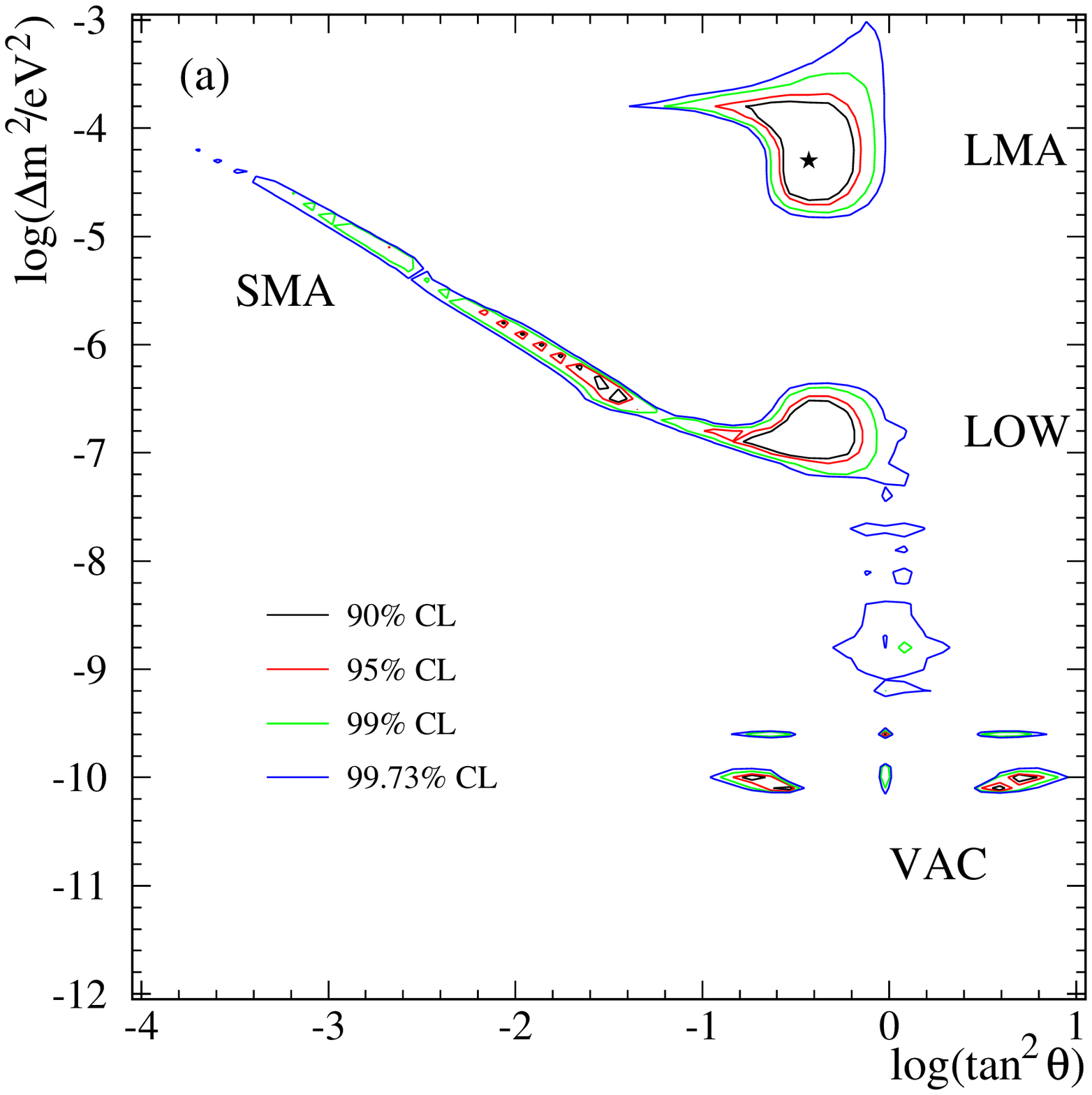}
\includegraphics[width=3.5in]{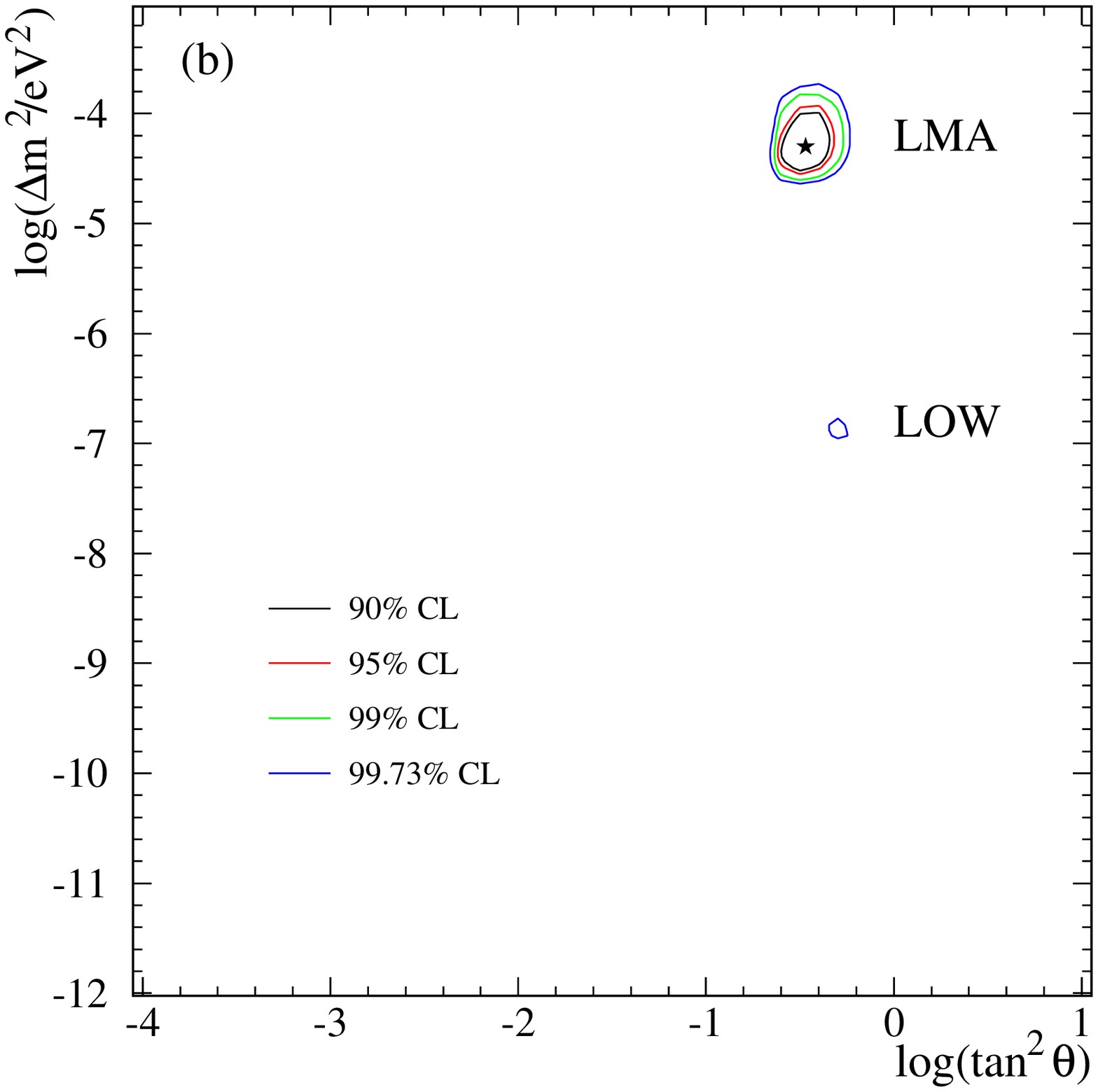}
\caption{\label{fig:MSWsno}
Allowed regions of the MSW plane determined  by a $\chi^2$ fit to (a) 
SNO day and night energy spectra and (b) with additional
experimental and solar model data.  The star indicates the best fit.
See text for details.} 
\end{center}
\end{figure*}

SNO's day and night energy spectra have
also been used to produce MSW exclusion plots and limits on neutrino
flavor mixing parameters~\cite{bib:snoprl2}.  MSW oscillation models
between two active flavors were fit to the data.  For simplicity, only
the energy spectra were used in the fit, and the radial $R$ and
direction $\cos\theta_{\odot}$ information was omitted.  This
procedure preserves most of the ability to discriminate between
oscillation solutions.  A model was constructed for the expected
number of counts in each energy bin by combining the neutrino spectrum, 
the survival probability, and the cross sections
with SNO's response functions.

There are 3 free parameters in the fit: the total $^8$B flux
$\phi_{B}$, the difference $\Delta m^2$ between the squared masses of
the two neutrino mass eigenstates, and the mixing angle $\theta$. The
flux of higher energy neutrinos from the solar \emph{hep} reaction was
fixed at $9.3 \times 10^{3}$~cm$^{-2}$~s$^{-1}$~\cite{bib:bpb}.
Contours were generated in $\Delta m^2$ and $\tan^2 \theta$ for
$\Delta \chi^2(c.l.) = 4.61~(90\%), 5.99~(95\%), 9.21~(99\%)$, and
$11.83~(99.73\%)$.  

Fig.~\ref{fig:MSWsno}(a) shows allowed mixing
parameter regions using only SNO data with no additional experimental
constraints or inputs from solar models.  By including flux
information from the Cl and Ga
experiments, the day and night spectra from the SK
experiment, along with solar model predictions for
the more robust $pp$, $pep$ and $^7$Be neutrino
fluxes~\cite{bib:bpb}, the contours shown in Fig. \ref{fig:MSWsno}(b)
were produced.  This global analysis strongly favors the Large Mixing
Angle (LMA) region, and $\tan^2 \theta$
values $<1$.  While the absolute chi-squared per degree of freedom is
not particularly large for the LOW solution, the difference between
chi-squared values still reflects the extent to which one region of
MSW parameter space is favored compared to another.  Repeating the
global analysis using the total SNO energy spectrum instead of
separate day and night spectra gives nearly identical results.

These results have been analyzed by many other authors.
In general, only the LMA, LOW and in some cases the VAC regions remain at the 3 sigma level,
with the LMA region strongly favored. The authors of reference~\cite{sterile} have shown an
upper limit of about 33\% for the possible sterile neutrino flux from the sun.

\section{Future Measurements}

All of the experiments with data to date plan to continue with solar neutrino measurements.
Kamiokande has been converted to the Kamland experiment by the addition
of liquid scintillator and will have a lower threshold. The Chlorine, GNO and Sage experiments 
will add to the accuracy of their measurements by further counting and will
improve upon the determination of the low energy region of the solar neutrino spectrum. Super-Kamiokande
is in the process of re-filling the detector after reinstalling the remaining phototubes following
the major implosion last year.
The community was shocked by the accident that befell the Super-Kamiokande 
experiment in November, 2001 and is very hopeful that the efforts now in progress will re-establish the
capability of this great experiment as soon as possible.

The SNO experimental plan calls for three phases where
different techniques are employed for the detection of neutrons from the NC 
reaction on deuterium. The NC reaction has a threshold of 2.2 MeV and is observed through the detection 
of neutrons by the three different techniques.
During the first phase, with pure heavy water, neutrons were observed 
through the Cerenkov light produced when neutrons are captured in deuterium, 
producing 6.25 MeV gammas. In this phase, the capture probability for such 
neutrons is about 25\% and the Cerenkov light is relatively close to the 
threshold of about 5 MeV electron energy, imposed by radioactivity in the 
detector. For the second phase, started in June, 2001, about 2.5 tonnes of NaCl were 
added to the heavy water and neutron detection is enhanced through capture 
on Cl, with about 8.6 MeV gamma energy release and about 83\% capture efficiency. 
For the third phase, due to begin in 2003, the salt will be removed and an array of $^{3}He$-
filled proportional counters will be installed to provide direct detection of 
neutrons with a capture efficiency of about 45\%. With the added sensitivity to neutrino types
other than electron neutrinos via the NC reaction, the accuracy for the determination of neutrino
flavor change will be increased significantly.

There are many other experiments planned in the future to provide measurements of solar neutrinos with
lower energy thresholds in real time. The Kamland experiment will report results soon and could have the
capability to observe $^7$Be neutrinos. The BOREXINO experiment should also have the capability for observing
$^7$Be neutrinos and could have lower radioactive backgrounds in that region. A comprehensive summary of other
planned low-energy solar neutrino measurements has been presented by S. Schonert at the Neutrino 2002 conference.

Many authors have analyzed the future capabilities of the existing experiments for restricting the
allowed regions of parameter space in future. A major restriction on oscillation parameters for active neutrinos
could be provided by the results from the Kamland experiment for reactor neutrinos if the correct solution is LMA.
When these results are combined with the SNO results, significant restrictions can be placed on sterile neutrinos 
without reliance on solar model calculations~\cite{sterile}. Initial solar fluxes of electron neutrinos
can also be determined with
high accuracy through the combination of these experimental results.

In summary, the set of solar neutrino experiments to date have provided a clear picture of neutrino
properties and definitive tests of solar models. The future measurements show excellent promise
for a fuller understanding of neutrino and solar properties as these and other experiments proceed.

\end{document}